\documentclass[12pt,preprint]{aastex62}

\newcommand{\kms}{\rm km~s^{-1}}

\usepackage{natbib}
\usepackage{multirow}
\usepackage{graphicx}	
\usepackage{amsmath}	
\usepackage{amssymb}	

\begin{document}

\title{Spectroscopic Tomography:  A First Weak Lensing Detection Using Spectroscopic Redshifts Only} 
\shorttitle{spectroscopic tomography}

\author{Ian Dell'Antonio}
\affil{Department of Physics, Brown University, Box 1843, Providence, RI 02912, USA}

\author{Jubee Sohn} \author{Margaret J. Geller}
\affil{Smithsonian Astrophysical Observatory, 60 Garden Street, Cambridge, MA 02138, USA}

\author{Jacqueline McCleary},
\affil{Jet Propulsion Laboratory, California Institute of Technology, 4800 Oak Grove Drive, Pasadena, CA 91109, USA
}
\author{Anja von der Linden} 
\affil{Department of Physics and Astronomy, Stony Brook University, Stony Brook, NY 11794, USA}

\begin{abstract}
We describe the first spectroscopic tomographic (spectrotomographic)  
 weak lensing measurement for a galaxy cluster based only on background galaxies with spectroscopically determined redshifts. 
We use the massive cluster A2029 to demonstrate the power of combining spectroscopy and lensing to obtain accurate masses and
 to overcome biases from contamination and photometric redshift errors. 
We detect the shear signal from the cluster at $>3.9 \sigma$ significance. 
The shear signal scales with source redshift in a way that is consistent with the angular diameter distance ratio variation in a $\Lambda$CDM Universe. 
Furthermore, the amplitude of the measured signal is consistent with the X-ray mass. 
Upcoming spectroscopic instruments such as the Prime Focus Spectrograph on Subaru will permit spectrotomographic weak lensing measurements with S/N 
 comparable to current photometric-redshift-based weak lensing measurements for hundreds of galaxy clusters. 
Thus, spectrotomography may enable sensitive cosmological constraints that complement and are independent of other measurement techniques. 
\end{abstract}

\section{INTRODUCTION}\label{sec:Introduction}

In the 30 years since its first demonstration \citep{Tyson90, Fort88},  
 weak gravitational lensing has become an important tool for measuring the distribution of mass in the universe on scales ranging from galaxies
 \citep{Brainerd96, Fischer00}
 to large scale structure \citep{Abbott18, Hildebrandt16}.
Weak lensing has also provided increasingly precise measurements of the masses of clusters of galaxies 
 (e.g. \citealp{Dahle02, Okabe10, Medezinski10, Umetsu14, vonderLinden14, Hoekstra15, McClintock19}). 
A recent compilation of weak lensing observations \citep{Sereno15}
 includes 485 unique systems. For ground-based weak lensing observations, 
 these studies have typically focused on clusters with $0.2<z<0.5$, 
 in the optimal redshift range for maximizing the shear and signal to noise within a small ($\sim 10\arcmin$) region. 

The availability of cameras with wider fields of view have enabled investigations of clusters at redshift $z <  0.1$ with comparable signal to noise.  
However, study of these systems still lags behind the study of higher redshift systems.  
For example, only $\sim 5\%$ (27 of 485) clusters used in \citet{Sereno15} are at $z < 0.1$.
Low redshift clusters are, however, important because they are easier to observe in detail with multiple techniques. 
Many now have high-quality X-ray data and spectroscopic measurements of hundreds or even thousands of galaxies (e.g. \citealp{Sohn19a, Sohn19b}). 

So far, weak lensing investigations of the lowest redshift systems, including the famous Coma Cluster \citep{Kubo07, Okabe14}, 
 use redshifts primarily to restrict the contamination of the weakly lensed source galaxies by unrecognized cluster members \citep{Okabe14}. 
However, spectroscopy can play a much greater role in cluster mass measurements 
 because biases in photometric redshift measurements are a dominant source of uncertainty in the weak lensing mass calibration \citep{vonderLinden14}. 
On the basis of a Bayesian analysis of photometric redshifts for individual weakly lensed sources,
 \citet{Applegate14} make a strong case for the power of combining redshift surveys with weak lensing observations 
 to suppress several of the systematics that impact cluster weak lensing results.

Here we demonstrate that combining weak lensing with deep, dense spectroscopy holds promise for direct evaluation and
 possible resolution of many subtle issues that affect the derivation of weak lensing cluster masses. 
As a first example, we investigate the weak lensing signal for the massive system Abell 2029 ($z = 0.078$)
 based on a large, complete set of spectroscopic redshifts that includes background galaxies. 
The low cluster redshift makes a sufficient number of weakly lensed sources accessible to
 the Hectospec instrument \citep{Fabricant05} mounted on the MMT 6.5 meter telescope. 
This redshift survey provides a basis for direct calibration of the weak lensing signal, 
 for evaluation of a weak lensing signal undiluted by cluster members, 
 and for direct detection of superimposed background structures. 

Because the majority of weak lensing studies make use only of photometric redshifts, 
 there are increasingly sophisticated photometric approaches to the problems of contamination 
 (both of the background galaxy sample and of the mass from superposition) and calibration of the photometric redshifts 
 (e.g. \citealp{Medezinski07, Mandelbaum10, Applegate14, Hoekstra15, Umetsu16}). 
These techniques have led to steadily improving limits on systematic biases.
Oddly, there has been no previous attempt to use a nearby cluster to measure
 the weak lensing signal based solely on a dense spectroscopic survey.
 
Our dense redshift survey of A2029 includes more than 5000 galaxies within $40\arcmin$ of the cluster center; 
 the redshift survey is essentially complete to $r = 20.5$. 
For the first time, we detect the cluster weak lensing signal. 
Although the uncertainty in the mass calibration is large given the relatively smaller number of background galaxies with spectroscopic redshifts, 
 the cluster mass we derive depends only on known geometric factors and does not depend on any external distance calibration. 
Furthermore, the set of weakly lensed sources contains no cluster members. 
A2029 is a very rich system; the redshift survey robustly identifies cluster members. 
We use these members to make two tests of potential weak lensing systematics. 

The A2029 system provides a measure of the weak lensing-induced ellipticity as a function of redshift (the ``tomographic" shear signal). 
We base the measurement exclusively on sources with spectroscopically measured redshifts. 
Although the uncertainties in the measure are large, 
 the observed behavior is consistent with previously published mass estimates for A2029 \citep{Sohn19a}.

Several investigators have suggested a cosmological test 
 based on the reduced shear as a function of redshift for a modest well-chosen set of clusters
 (\citealp{Medezinski11}; Figure 1 of \citealp{Applegate14}). 
The A2029 experiment demonstrates the feasibility of a cosmological test based on much larger, deeper, redshift surveys of well chosen 
 sets of clusters carried out with future instruments like Subaru/Prime Focus Spectrograph (PFS, \citealp{Tamura16})
 or GMT/GMACS \citep{DePoy18}. 

In Section~\ref{sec:Phot}, we describe the photometric data acquisition and reduction used to produce the object catalog and to derive the ellipticities of the objects. 
In Section~\ref{sec:Spectra}, we describe the spectroscopic data. 
In Section~\ref{sec:Shear}, we describe the ellipticity measurements for the objects with spectroscopic redshifts. 
In Section~\ref{sec:Profile_Tomography}, we derive the ellipticity profile versus projected radius and the tomographic measurements.  
Finally, in Section~\ref{sec:Discussion}, we discuss the spectrotomographic detection and the prospects for exploiting 
 the combination of weak lensing shear and spectroscopic redshifts with upcoming massively multiplexed fiber instruments to make sensitive cosmological predictions.

\section{Photometric Data analysis and catalog generation}\label{sec:Phot} 

\subsection{Photometric Data and Processing}\label{subsec:imagedata}

Photometric observations were taken with the Dark Energy Camera (DECam) at the Cerro Tololo Inter-American Observatory's 4-meter telescope. 
The DECam imager consists of 62 2048 $\times$ 4096 pixel science CCDs (60 of which are currently operational) arranged in a hexagon.  
DECam images cover 2.2 square degrees at a 0\farcs265/pixel scale \citep{DePoy08, Flaugher15}. 

The A2029 data for this project were obtained during four observing runs (04/26-28/2013, 06/11-16/2013, 08/29-31/2014, 03/30-04/02/2014).  
The data were obtained as part of two separate observing programs: a dedicated campaign by Jacqueline McCleary to look for cluster substructure, 
 and a DECam program by Anja von der Linden to obtain scaling relations for cluster cosmology. 
The observations cover  five filter bands ($ugriz$), to allow photometric redshift calculation.  
For our analysis, we only make use of data in the filter band with the best seeing ($i$). 
Observations were taken as sequences of dithered exposures (each typically 120s or 300s).  
Only exposures taken in seeing better than $1\arcsec$ were used in the shape analysis.  
The final exposure time for these ``good-seeing''$i$-band observations was 4920 seconds.

Instrumental signatures were removed using the DECam Community Pipeline (CP). 
The CP performs: bias calibration; crosstalk; masking and interpolation over saturated and bad pixels; 
 CCD non-linearity and the flat field gain calibration; fringe pattern subtraction; astrometric calibration; 
 single exposure cosmic ray masking; characterization of photometric quality; sky pattern removal; and illumination correction.  
For full descriptions of the DECam pipeline processing system, see chapter 4 of the NOAO Data Handbook~\citep{NOAODHB2.2}.

\subsection{Image Stacking and Calibration}\label{subsec:photometry}

We base our analysis of the deep lensing-quality image in \citet{MDvdL_2018}. 
For convenience, we summarize the steps leading to this image. 

\citet{MDvdL_2018} create the stacked image using a weighted stacking algorithm.  
They calculate weights for the individual exposures based on their photometric depth, 
 obtained by matching instrumental magnitudes of unsaturated stars to those in the Sloan Digital Sky Survey (SDSS) Data Release (DR) 12 catalog \citep{Alam15}.

They resampled individual scaled exposures onto a common astrometric grid and combined them into a stacked image using {\sc SWarp}~ \citep{Bertin10}
 which also performs the sky subtraction on the individual exposures allowing for a uniform sky even in the presence of uneven weights 
 due to gaps and defects in individual exposures.  
For the projection step, \citet{MDvdL_2018} use \texttt{Lanczos3} kernel, chosen because of its robust noise-conservation properties and computational efficiency. 

To produce the final image, \citet{MDvdL_2018} use the clipped mean extension to {\sc SWarp}.
This approach is exceptionally stable to a wide range of artifacts in individual frames and
  produces a stacked image where the point-spread function (PSF) is a linear combination of the single frame PSFs \citep{Gruen14}. 
Thus the clipped mean stacks are well-suited for weak lensing shape measurement.

 The image stacking process is described in \citet{MDvdL_2018} and is based on the process developed by the Weighing the Giants project
\citep{vonderLinden14}. \citet{MDvdL_2018} test the quality of the image stack by selecting stars in the individual exposures via their FWHM and surface brightness profile, and use these to estimate the scale of the PSF variations by fitting spatial averages of the stellar ellipticity components $e_1$ and $e_2$ across the image.  

Residual PSF shape variations from the fit are $\ll 0.5$\%  
 on angular scales larger than $1\arcmin$ over the central $1.5^{\circ}$ of the field.   Because this is much smaller than the mean ellipticity signal we measure for the galaxies in \ref{sec:Shear}, we conclude that PSF measurement errors over the angular region we consider (within 23$^{\prime}$ of the cluster center) do not contribute significantly to the shear measurement errors. 
In Section \ref{sec:Shear} we use the lensing quality image for the weak lensing analysis.

\section{Spectroscopic Data}
\label{sec:Spectra}

\citet{Sohn19a} carried out a dense and complete spectroscopic survey of A2029.  
Within $30^{\prime}$ of the cluster center, the redshift survey is 90\% complete to $r = 20.5$.
We use the redshifts to identify cluster members and foreground/background galaxies. 
The background redshifts are the basis for the spectroscopic tomography. 

To construct the complete spectroscopic survey, 
 we first collected redshifts of  bright A2029 galaxies from the SDSS. 
 The SDSS sample primarily {\bf consists} of  the spectra for galaxies brighter than $r = 17.77$ through $3\arcsec$ fibers \citep{2002AJ....124.1810S}.   
There are 681 objects with SDSS spectra within $40\arcmin$ of the cluster center. 
The typical redshift uncertainty for SDSS spectra is $13~\kms$. 

We also collected redshifts of 41 A2029 galaxies from the literature (e.g. \citealp{Bower88, Sohn17}), 
 accessed through the NASA/IPAC Extragalactic Database (NED).
\citet{Tyler13} measured redshifts of A2029 galaxies using Hectospec mounted on MMT 6.5m telescope, 
 a fiber-fed spectrograph that obtains $\sim 250$ redshifts over an $\sim 1~{\rm deg}^{2}$ field of view \citep{Fabricant05}. 
We collected 1362 redshifts from \citet{Tyler13} through the MMT archive \footnote{http://oirsa.cfa.harvard.edu/archive/search/}. 
 
We then conducted a denser, more complete redshift survey of A2029 using MMT/Hectospec \citep{Sohn17, Sohn19a}. 
The survey targets are galaxies with $r_{\rm petro} = 21.3$ and $r_{\rm fiber} = 22$ in the SDSS photometric galaxy catalog. 
We imposed the fiber magnitude selection to exclude low surface brightness galaxies from the target list. 
There is no color selection for survey targets.  

For the Hectospec observations , we used the 270 line mm$^{-1}$ grating also used by \citet{Tyler13}. 
The resulting spectra have a resolution of 6.2~{\rm \AA} covering a wavelength range $3700 < \lambda < 9150$ {\rm \AA}.  
We observed each field with three exposures of 1200 seconds to allow for cosmic ray removal. 

We used the HSRED v2.0 IDL package to reduce the Hectospec spectra. 
We derive the redshifts using the RVSAO \citep{Kurtz98} package that 
 cross-correlates Hectospec spectra and a set of template spectra prepared for this purpose \citep{Fabricant05}.
We visually inspected the reduced spectra and divided the results into three categories:
 `Q' indicates a high-quality redshift, `?' an ambiguous result, and `X' denotes a poor quality redshift. 
We use only the 2890 high-quality redshifts. 
The typical redshift uncertainty for Hectospec spectra is $\sim 29~\kms$. 

A2029 is one of the best-sampled galaxy cluster fields. 
Within $30^{\prime}$ of the cluster center defined by the brightest cluster galaxy (BCG),
 IC1101, there are 3568 objects with measured redshifts. 
The spectroscopic completeness is $\sim90\%$ to $r < 20.5$ within $R_{cl} < 30\arcmin$ ($\sim 67\%$ to $r < 21.3$).
The spectroscopic survey is uniform within $R_{cl} < 30 \arcmin$ (see Figure 1 in \citealp{Sohn19a}). 

\begin{figure}
\centering
\includegraphics[scale=0.4]{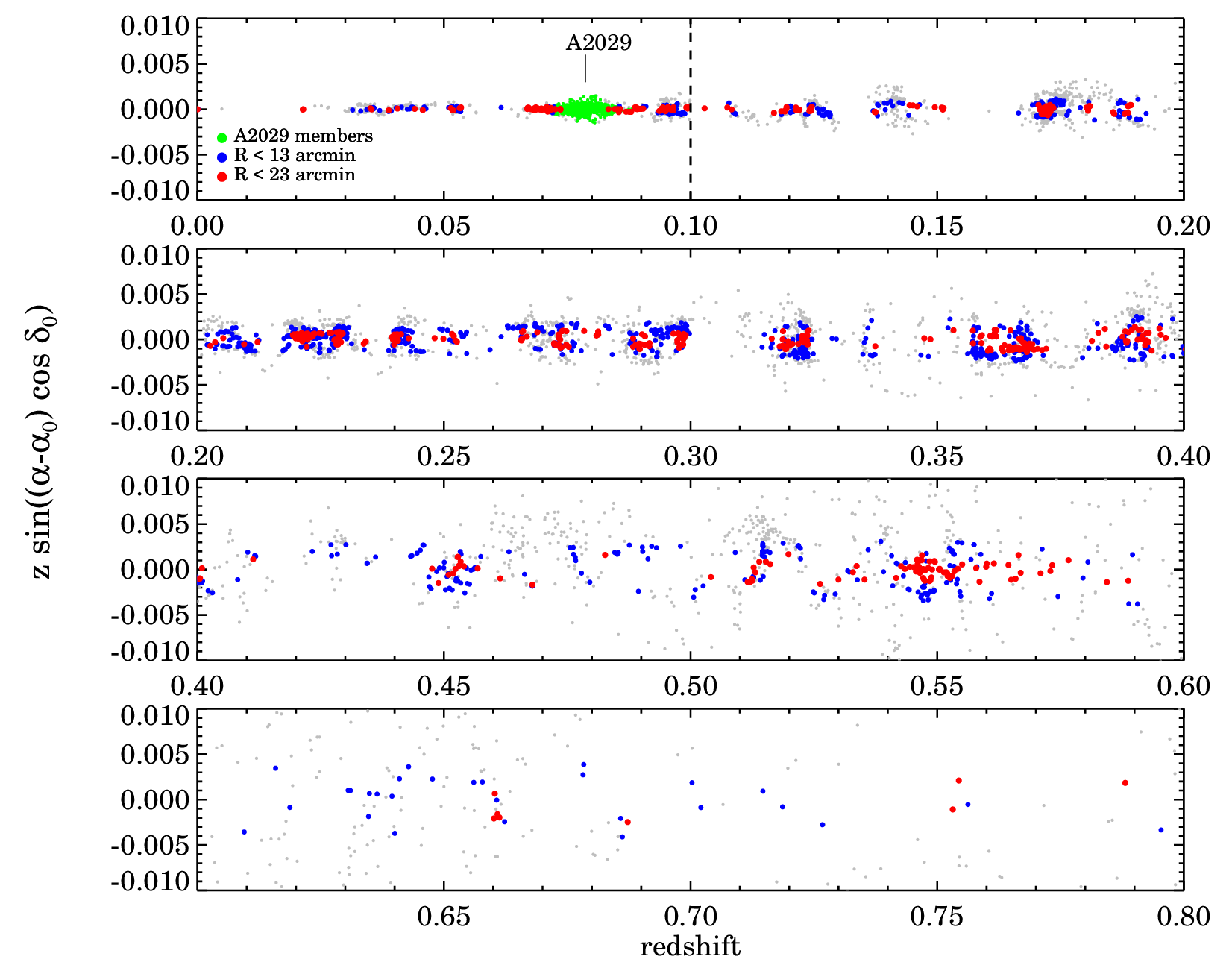}
\caption{Cone diagram of the A2029 field projected in the R.A. direction. 
Gray points are the galaxies with a spectroscopic redshift.  
Green points are spectroscopically identified members of A2029. 
 Blue and red points are galaxies within $R < 13\arcmin$ and with $13\arcmin < R < 23\arcmin$, respectively. 
We use these galaxies at $z > 0.1$ for the weak lensing analyses. }
\label{fig:A2029cone}
\end{figure}

Figure \ref{fig:A2029cone} shows a cone diagram for the entire redshift survey. 
The finger corresponding to A2029 is obvious at mean redshift of 0.078. 
The voids and filaments characteristic of the large-scale structure of the universe are evident particularly in the background. 
There is also foreground structure which we use below to test the robustness of the lensing signal.  
We highlight the two radial ranges where we explore the weak lensing signal: 
 the blue points highlight objects within the approximate virial radius, $R = 23 \arcmin$. 
We highlight the dense central region of the cluster (R $ < 13\arcmin$) in red.
 
We use the total redshift survey to  identify cluster members and foreground/background objects.
We identify cluster members using the caustic technique \citep{Diaferio97, Diaferio99, Serra13}, 
 a non-parametric characterization of the cluster boundary. 
The caustic technique estimates the boundary of clusters in the phase space, 
 the relative velocity difference with respect to the cluster mean velocity as a function of cluster-centric distance. We identify
galaxies within the cluster boundary phase space 
as cluster members (see Figure 4 in \citealp{Sohn19a}). 

Table \ref{tabstat} summarizes the various subsets of the spectroscopic sample relevant for the spectroscopic tomography. 
The table includes the number of galaxies with both a redshift and an ellipticity measurement that are cluster members,
 foreground galaxies, or background objects
 within two projected radii ($13\arcmin$ and $23\arcmin$) of the cluster center defined either by the weak lensing map or by the position of the BCG. 
We use the samples of background galaxies with redshifts and ellipticities for the spectrotomographic measurements.

\begin{deluxetable}{lccccc}
\label{tabstat}
\tablecolumns{6}
\tabletypesize{\scriptsize}
\tablecaption{A2029 Redshift Samples}
\tablehead{
\multirow{2}{*}{Type}	& \multicolumn{2}{c}{WL Center} & &  \multicolumn{2}{c}{BCG} \\
										& $R < 13'$ & $R < 23'$ & & $R < 13'$ & $R < 23'$}
\startdata
Spec. objs											& 859	& 2252 &	&	860	&	2256	\\ 
Spec. objs with shear measurements	& 854	& 2225 &	&	855	&	2228	\\
Spec. member									& 339	&	595	&	&	338	&	597	\\ 
Spec. foreground								& 46		&	100	&	&	46		&	99		\\  
Spec. background								& 474	& 1517	&	&	466	& 1519	\\
Spec. background, $z > 0.10$			& 433	& 1431	&	&	426	& 1432	\\
\enddata
\end{deluxetable}

\section{Weak Lensing Shear Measurements}
\label{sec:Shear}

Weak lensing shear measurements require fitting for and {\bf removing} 
the effect of the PSF on the galaxy shapes.  

Several methods have been developed to extract the PSF-corrected ellipticity components 
 ({\it e.g.} KSB \citep{Kaiser95}, Bernstein \& Jarvis \citep{Bernstein02}).

For the galaxy shape measurement, we use the well-established ``Regaussianization'' technique of Hirata, Seljak and Mandelbaum (HSM) \citep{2003MNRAS.343..459H, Hirata04, 2005MNRAS.361.1287M}.  

We use the implementation of regaussianization included in the LSST Software data management system (the Rubin Observatory LSST Science Pipelines).  Specifically, we use the 13.0 release of the LSST Science Pipeline, which implements regaussianization using the GalSim implementation \citep{2015A&C....10..121R}.
 
 For PSF estimation and the modeling of its spatial variation, we use the default PCA-based modeling algorithm (cf. \citep{2004astro.ph.12234J}, \citep{2006JCAP...02..001J}) built into the v13.0 release of the science pipeline.  Stars for the fit were selected via their ``extendedness", based on the difference of aperture magnitudes over multiple apertures (this serves to separate stars from galaxies) and maximum and minimum flux (to avoid saturated stars, as well as stars detected with S/N less than 50).

This implementation of regaussianization is similar to that used for analyzing the Subaru Hyper Suprime-Cam (HSC) Subaru Strategic Program (SSP) survey \citep{Mandelbaum18, Hikage19} 
  but uses an older implementation of the LSST Science Pipelines.  The PSF modeling is different, as more recent version of the science pipelines use PSFEx \citep{2013ascl.soft01001B} as the default PSF modeling method.  In addition, because the galaxies with spectroscopic measurements are all well-resolved and quite bright, we do not apply any filtering to the shape catalog. 
To access these algorithms, we use the {\it obs\_file} package \footnote{https://github.com/SimonKrughoff/obs\_file}
 which allows running of the core LSST analysis packages on individual stacked images (and which is compatible with the V13.0 release of the DM Science Pipelines).  
We use the stacked image produced in Section \ref{sec:Phot} as the input for the shape measurement. 
We cross-correlate the output catalog with the spectroscopic catalog.   
The imaging catalog detects about 1.2 million objects (mostly galaxies in the A2029 field with significance greater than $5\sigma$;  
 almost all of these galaxies are fainter than the spectroscopic limit). 
Because source crowding is an issue, we use the built-in deblending algorithm in the DM Science Pipelines
 to separate sources and to measure shapes for the deblended objects.  
We select only objects without deblended sub-objects ($N_{child}=0$).  
This selection could bias the photometry for the largest galaxies in our sample because the deblending algorithm tends to oversplit split large complex galaxies.
However, we do not rely on this photometry when matching our sources to the spectroscopic sample; 
 we rely only on centroid positions. 
We use the ellipticity components ($e_1$, $e_2$) from Regaussianization to calculate the tangential ellipticity profile and the tomographic signal for A2029.

We filter the photometric catalog to include only objects with $i<23.5$ before matching to the spectroscopic catalog.  This reduces the source density to eliminate ambiguities in which multiple objects in the shape catalog match a spectroscopic detection.

Because of differences in deblending close neighbors and source multiplicity, 
 not every spectroscopically measured object has a shape in the Regaussianization catalog. 
 
Table \ref{tabstat} lists 
 the relevant numbers of spectroscopic objects and spectroscopic objects with shapes 
 for each subsample we analyze. 
All of the objects without shapes are either stars ($\sim 53$\%), 
 cluster members ($\sim 12$\%) or foreground objects ($\sim 35$\%).

\citet{MDvdL_2018} demonstrate a statistically significant weak lensing signal $R = 23\arcmin$ of the lensing center.  
We thus restrict the subsequent analysis to galaxies within this radius.
The position of the highest lensing peak in \citet{MDvdL_2018} is offset by about $0.3\arcmin$ from the center of the BCG, IC1101. 
Table \ref{tabstat} gives the numbers of objects in samples centered on 
 the weak lensing peaks and on the BCG.  
In Section \ref{subsec:Qchecks}, we show that these two choices of the lensing center have minimal overall impact on the spectrotomographic signal.

\subsection{Weak Lensing Measurement}

\begin{figure}
\centering
\includegraphics[scale=0.4]{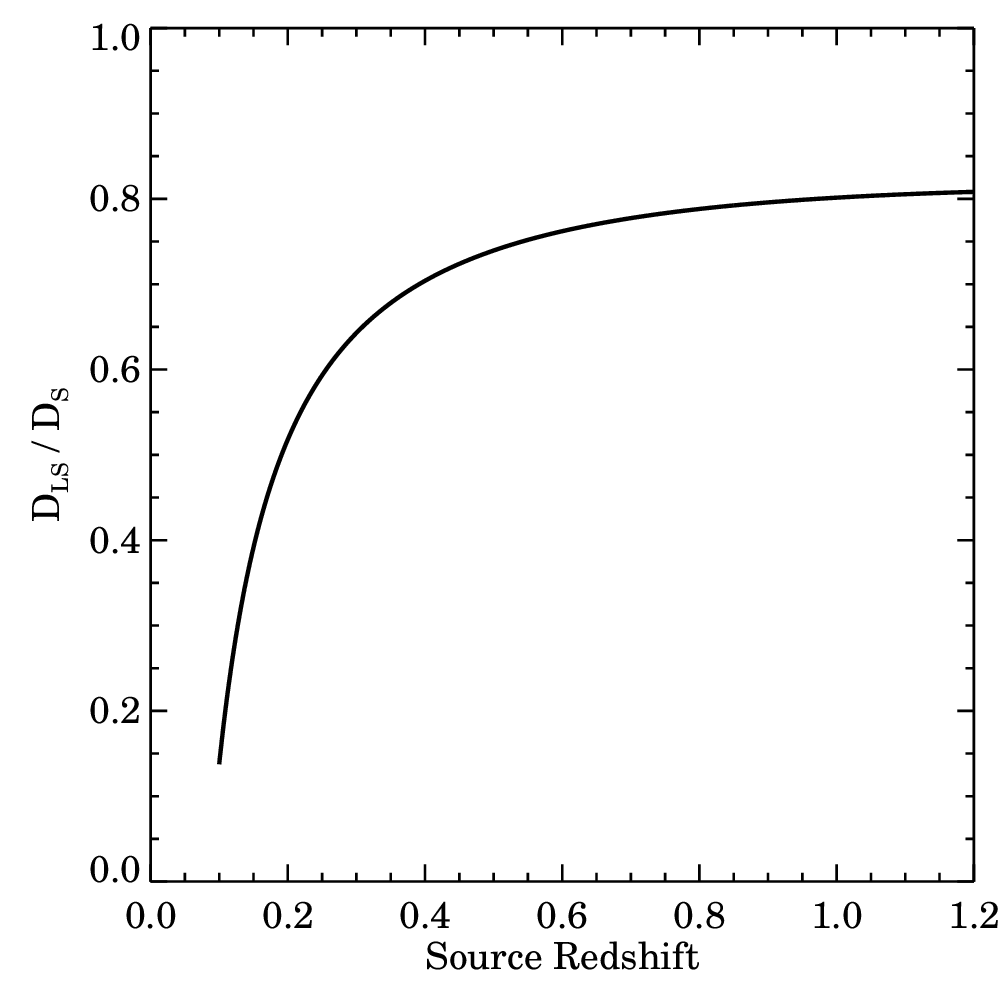}
\caption{The angular diameter distance ratio ${D_{LS}/D_S}$ 
 as a function of source redshift for galaxies behind A2029.  
Distances assume a $\Lambda$CDM cosmological model 
 with $\Omega_m=0.27$ and $\Omega_\Lambda=0.73$. }
\label{fig:dratio}
\end{figure}

In weak gravitational lensing, 
 the distortion of the images is described by the reduced shear $\boldsymbol{g} =\frac{\boldsymbol{\gamma}}{1-\kappa}$, 
 where $\boldsymbol{\gamma}$ represents the anisotropic shear induced in the galaxy images.  
The reduced shear is a function of the projected mass density of the lens.  
At fixed lens redshift, the reduced shear is a function of the source redshift through the angular diameter distance ratio $\frac{D_{\rm ls} }{D_{\rm s}}$, 
 which impresses a characteristic pattern on the lensing signal as a function of source redshift.  
For a low redshift cluster like A2029, the tomographic lensing signal is a steep function of redshift for $z>z_{cluster}$ (Figure \ref{fig:dratio}).

The reduced shear is related to the measured {\it tangential ellipticity}.   For the case of Regaussinization, the average tangential ellipticity can be written in terms of the reduced shear and the variance in ellipticities or shape noise: 
\begin{equation}
<e_{\tan}> = -<(e_1 \cos(2\phi) + e_2 \sin(2\phi))> \simeq { 2g (1-R)} \label{eqn:etan}
\end{equation}
Where $R=\sigma_e^2 = (0.37)^2$. In this paper we measure the lensing signal via the mean tangential ellipticity, but where we overplot expectations for given cluster mass profiles, we convert shear to tangential ellipticity via Equation 1.

The variables $e_1$ and $e_2$ in Equation~\ref{eqn:etan} are 
 the polarization states of background galaxies with complex ellipticities $\boldsymbol{e}$, 
 measured in terms of the image axes $x$ and $y$; 
 $\phi$ is the azimuthal angle between the $x$-direction and the vector 
 connecting the position of the center of the cluster's mass to the position of the galaxy.

\subsection{Expected Signal Extent}

According to \citet{MDvdL_2018} the distortion effects of A2029 are significant within $R = 23\arcmin$. 
\citet{MDvdL_2018} selected background galaxies based on their photometric redshifts.  
They selected galaxies with photometric redshift probability $P (z  >0.19) > 0.8$
  based on BPZ template-fitting \citep{Benitez00}.  
Although this procedure largely selects background galaxies, 
 there is significant cross-contamination of the background by faint cluster members.  
However, the photometric sample still provides a robust extent of the shear signal. 

Figure 16 of \citep{MDvdL_2018} shows that tangential shear for the photo-z sample is detected 
 at high signal to noise (and with reduced shear $g > 0.01$) to a radius $R=23\arcmin$,
 approximately the virial radius for A2029 \citep{Sohn17}. 
Although that  sample contains some cluster members, 
 the large number of galaxies used in the photometric shear measurement leads to
 a higher S/N measurement of the lensing signal than 
 we can obtain for the spectroscopic measurement described below.  
For this reason, we use $R < 23\arcmin$ as the radial extent over which 
 we measure the signal for the spectroscopic sample. 
  
\subsection{Quality Checks}
\label{subsec:Qchecks}

Systematic tests on the residual PSF ellipticities in the lensing image were already carried out in \citet{MDvdL_2018}. 
We perform additional tests on the shape catalog to verify that systematic errors that could mimic a shear signal are inconsequential.    
Additionally, we  use  the ``B-mode" or curl-like component of the ellipticity tensor to verify that the signal has the characteristics of genuine lensing.

 Although absence of bias in the ``B-mode" lensing does not necessarily imply lack of bias in the ``E-mode" component, the a-priori expectation is that PSF modeling errors should not preferentially select one mode over the other.  Therefore, the ``B-mode" signal serves as an extra control on the validity of the tangential ellipticity signal.

\subsubsection{Ellipticity Component Distributions for the Photometric Sample}

\begin{figure}
\centering
\includegraphics[scale=1.0]{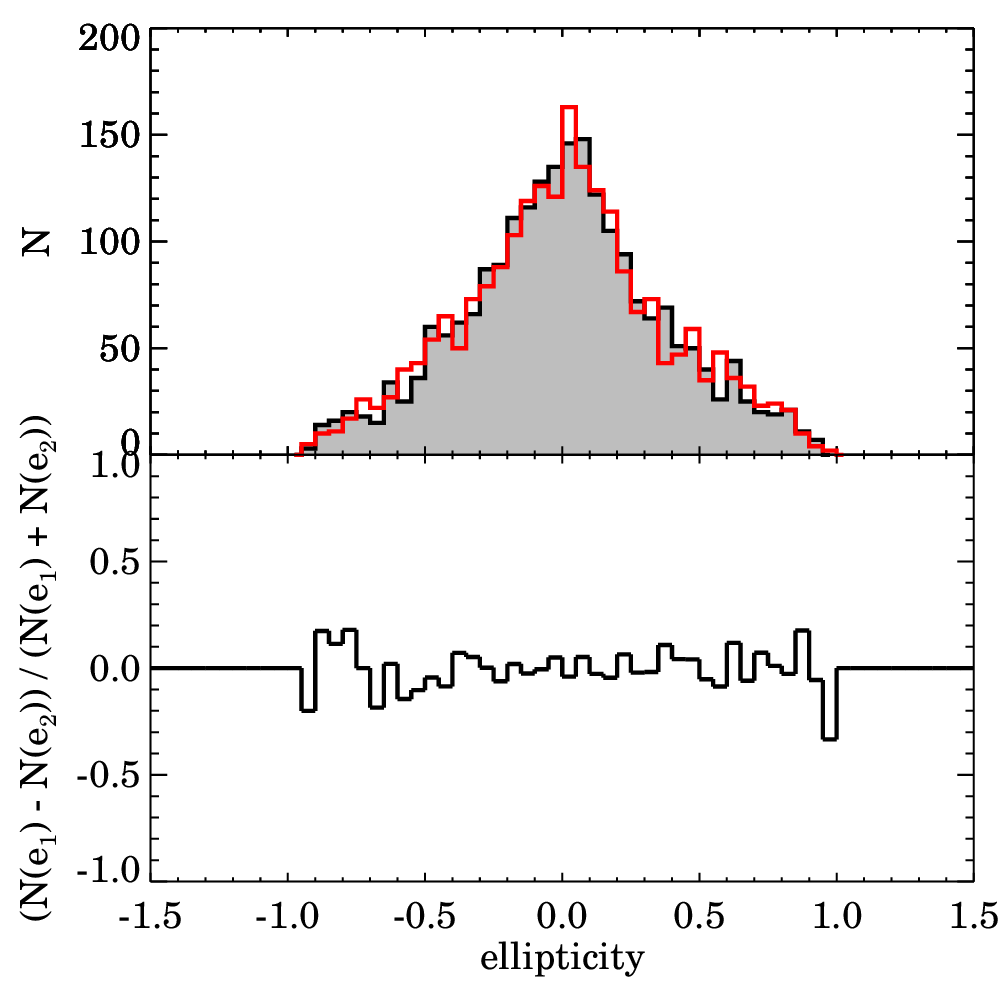}
\caption{(Top) Distribution of Regaussianization ellipticity estimates $e_1$ and $e_2$ for 
 the A2029 spectroscopically matched shape catalog within $R = 23\arcmin$. 
(Bottom) The differential ratio between $e_{1}$ and $e_{2}$ as a function of ellipticity. 
 The variation between the two distributions are consistent to within 1$\sigma$ assuming Poisson statistics.} 
\label{fig:A2029_shear_distribution}
\end{figure}

In the absence of systematic errors (and in the limit where $g \ll 1$) 
 the tangential ellipticity is an unbiased estimator of the reduced shear when averaged over galaxies.
As a test for systematic offsets in the galaxy shapes, 
 we plot the distribution of ellipticity components $e_{1}$ and $e_{2}$ 
 (corresponding to the ``plus" and ``cross" pattern ellipticities) 
 for galaxies in the Regaussianized catalog for the A2029 field.   
In the absence of systematic distortions, 
 the azimuthal symmetry of the vectors to the center of A2029 should result in 
 indistinguishable distributions $N (e_{1})$ and $N (e_{2})$ even in the presence of a lensing signal.  
The upper panel of Figure \ref{fig:A2029_shear_distribution} shows that
 the two distributions are similar: 
 the lower panel shows $(N(e_{1}) - N(e_{2})) / (N(e_{1}) + N(e_{2}))$ as a function of ellipticity.
The differences are within the 1$\sigma$ errors throughout the ellipticity range. 
The largest excursions at the extreme values of ellipticity in the bottom panel occur 
 because there is a negligible number of objects with that ellipticity.
 
\subsubsection{Tangential and ``B-mode" distribution for the Sample with Redshifts}

\begin{figure}
\centering
\includegraphics[width=0.9\columnwidth]{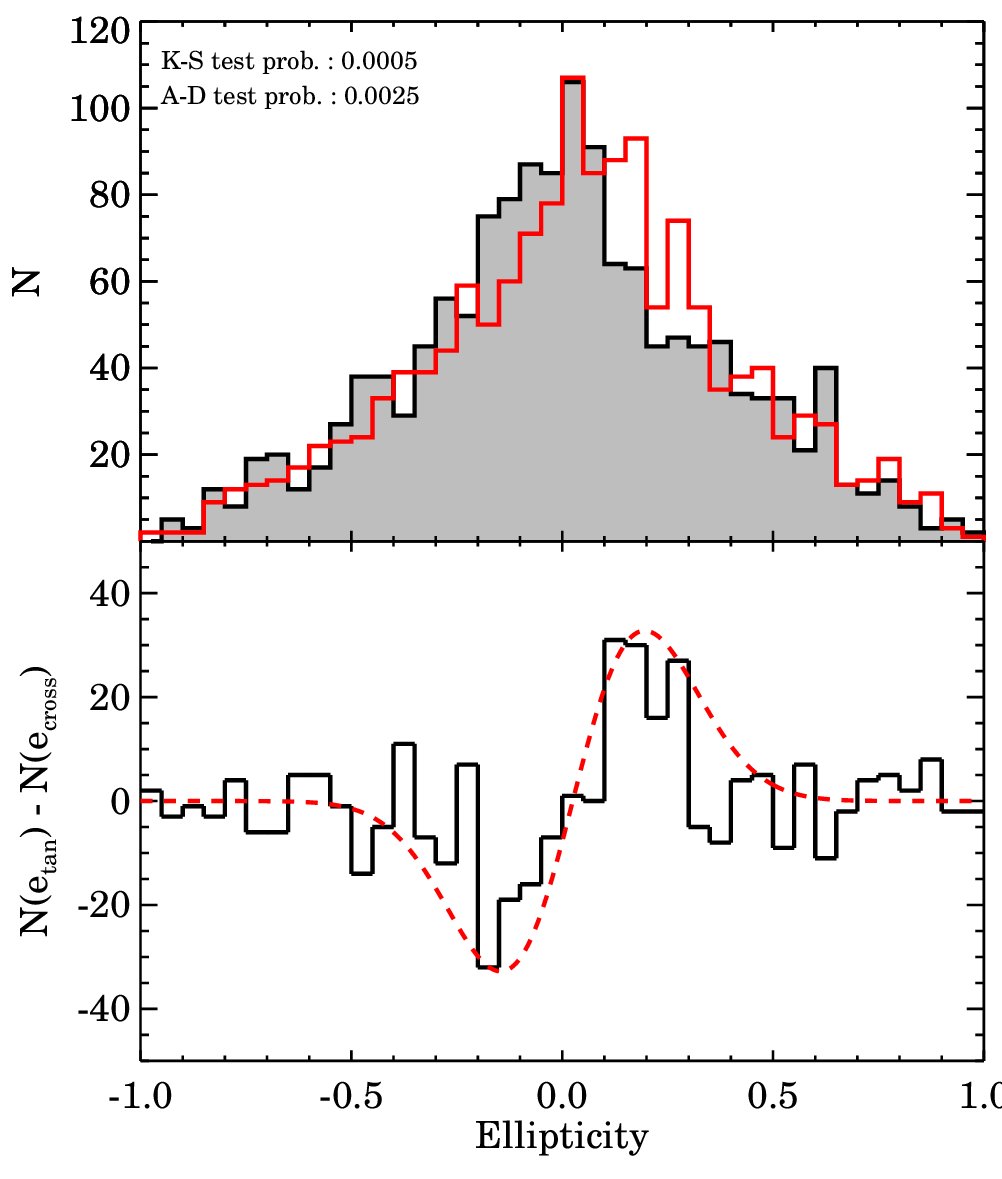} 
\caption{(Top) Distribution of tangential ellipticity $e_{tan}$ (red) and ``B-mode"  ellipticity $e_{cross}$(black)  for 
 the A2029 field galaxies within $R = 23\arcmin$ at $z > 0.1$ (A2029 background galaxies). 
(Bottom) The difference between the $e_{tan}$ and $e_{cross}$ distributions (in the upper panel) as a function of ellipticity. 
The superimposed curve shows the theoretical expectation for two normal distributions offset 
 by the mean shear signal observed for A2029. }
\label{fig:etanecrosshist}
\end{figure}

In the presence of lensing, 
 the distributions of $e_{cross}$ and e$_{tan}$ should differ with a systematic offset 
 towards non-zero values of $e_{tan}$.
The upper panel of Figure \ref{fig:etanecrosshist} shows histograms of the numbers of galaxies 
 with a given value of tangential ellipticity ($N_{etan}$) and 
 the number of galaxies with the same value of the cross (or B-mode) ellipticity ($N_{ecross}$). 
The lower panel of Figure \ref{fig:etanecrosshist} shows 
 the expected offset between the two distributions. 
We restrict the sample of objects to those within $R = 23 \arcmin$ and with $z > 0.1$.  
A Kolmogorov-Smirnov (KS) test demonstrates that the distributions in the upper panel are inconsistent with being drawn from 
 the same underlying distribution at a level $p = 0.0005$. 
The Anderson-Darling (AD) test gives a similar result. 
In essence the statistical significance given by the KS and AD tests is 
 our first measure of the significance of the spectroscopic tomographic signal. 
The offset between the two distributions is the mean lensing-induced tangential ellipticity. 
 
We can also examine the difference between the two distributions in Figure \ref{fig:etanecrosshist} 
 as a function of ellipticity to check for systematic bias in the spectroscopic sample selection.  
We can approximate the two offset distributions as Gaussians 
 with the width of the ellipticity distributions.  
The pattern of tangential ellipticities matches the theoretical expectation very closely.  

\subsubsection{Choice of Cluster Center}

Because we calculate the tangential and cross ellipticity components for each galaxy 
 with respect to the direction between the galaxy and the center of the cluster, 
 the choice of center could significantly affect  the measured tangential shear pattern (see Equation \ref{eqn:etan}).
To test the impact of the choice of center, we compute the spectrotomographic signal 
 using two different centers, the weak lensing peak \citet{MDvdL_2018} and the position of the BCG (see Table 1).  
  
\begin{figure}
\centering
\includegraphics[width=0.7\columnwidth]{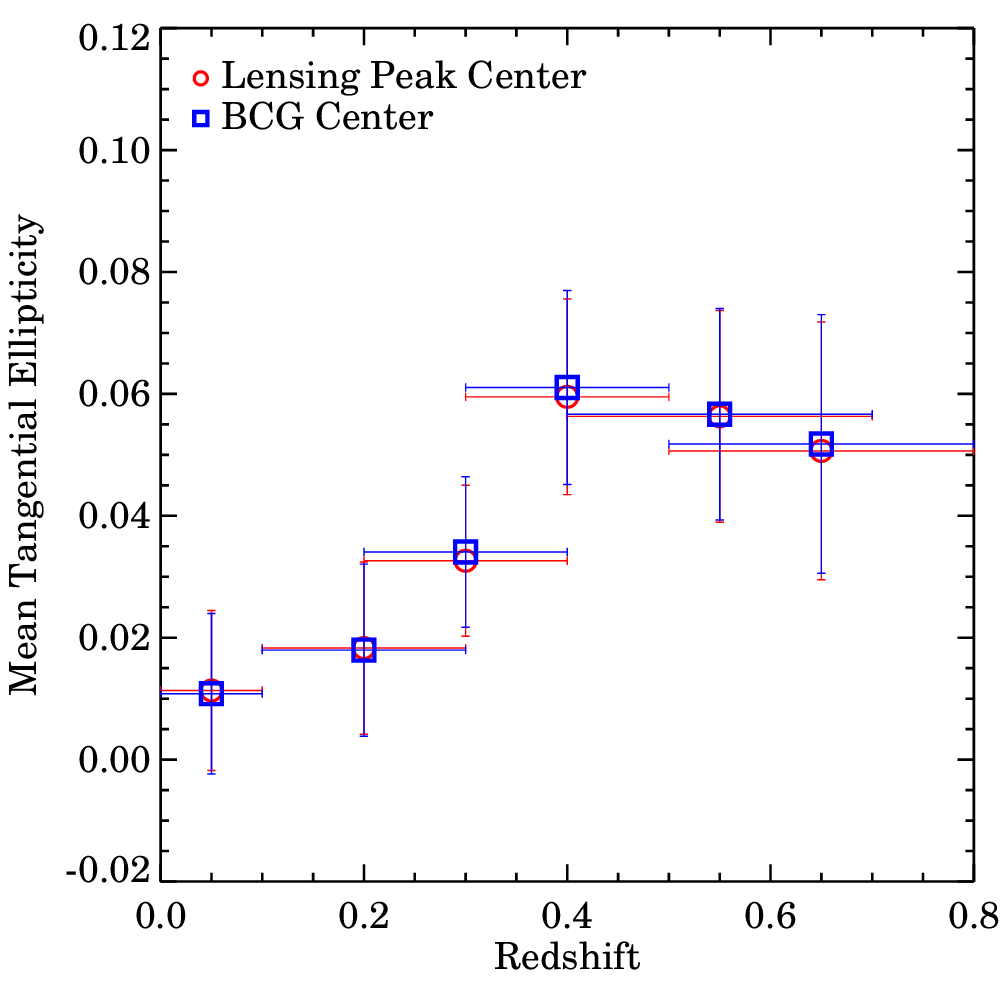}
\caption{The spectrotomographic signal for the $23 \arcmin$ sample with two different cluster centers.  
In all bins, the signals are within the $1\sigma$ uncertainties, 
 but the overall normalization of the signal depends very slightly on the choice of center. }
\label{fig:comparecenters}
\end{figure}

These two centers are within $0.3\arcmin$ of one another.
Figure \ref{fig:comparecenters} thus shows that the lensing signals are very similar.  
In each bin, the tangential ellipticities are consistent within the uncertainties and the overall normalization is also similar. 
On average, the lensing signal is $\sim 5\%$ lower 
measured with respect the center of IC1101 rather than with respect to the lensing signal.
It is perhaps not surprising that the weak lensing center maximizes the signal, but the effect is small. 
In the following analysis we use the weak lensing center.

\subsubsection{Cluster Members and Foreground Galaxies}

In principle, the shear and B-mode signals for the cluster galaxy sample test models of the intrinsic alignment of cluster galaxies (e.g. \citealp{Kiessling15, Huang18}). 
We use the caustic pattern of A2029 to select cluster members (see Section \ref{sec:Spectra}).
We can then compute the shear signal separately for cluster members and for unrelated foreground galaxies.  
There are 597 caustic-selected cluster members within a projected distance of $23\arcmin$ 
 of the center of A2029 (338 within $13\arcmin$). 
For these cluster galaxies the mean tangential ellipticity is consistent with zero to 
 within $1\sigma$ ($0.014\pm 0.017$) for the $13\arcmin$ sample. 
This result implies that intrinsic alignment is unimportant as suggested by other studies. 
The mean tangential ellipticity is larger in the larger radius sample; 
 it is marginally inconsistent with zero at $2.2\sigma$ ($0.030\pm 0.014$). 
This result is inconsistent with a model where intrinsic alignments result from 
 galaxy interactions \citep{Singh15} 
 and may indicate some unknown source of correlation in cluster galaxy orientations. 
The B-mode signal for cluster members is consistent with zero ($1\sigma$ negative) for 
 both 13 and $23\arcmin$ samples.

There are 101 foreground galaxies within $23 \arcmin$. 
These objects obviously should not be lensed. 
Both the  ``shear" signal and the B-mode from these galaxies is consistent with zero. 
The mean tangential ellipticity is $-0.014\pm 0.023$, and the B-mode signal is $0.025\pm 0.023$. 
The signal from the 698 galaxies within $23 \arcmin$ and including cluster members and the foreground is insignificant. 

The uncertainty in the measurement from this single cluster is too large 
 to distinguish among models which predict positive and negative tangential alignments. 
However with deep cluster samples of the future that could reach 3 magnitudes 
 deeper than the survey we use, the number of cluster member would increase by 
 a factor of 10 or more depending on the faint end slope of the luminosity function (LF, \citealp{Sohn17}).
The number we quote refers to a flat faint end; a steep faint end (e.g. \citealp{Agulli16})
 only increases the number. 
Uncertainties based on these large samples will be 
 a factor of $\sim4$ smaller thus providing powerful constraints on model predictions. 

\section{Shear Profiles and Tomography}\label{sec:Profile_Tomography}

\begin{figure*}
\centering
\includegraphics[scale=0.6]{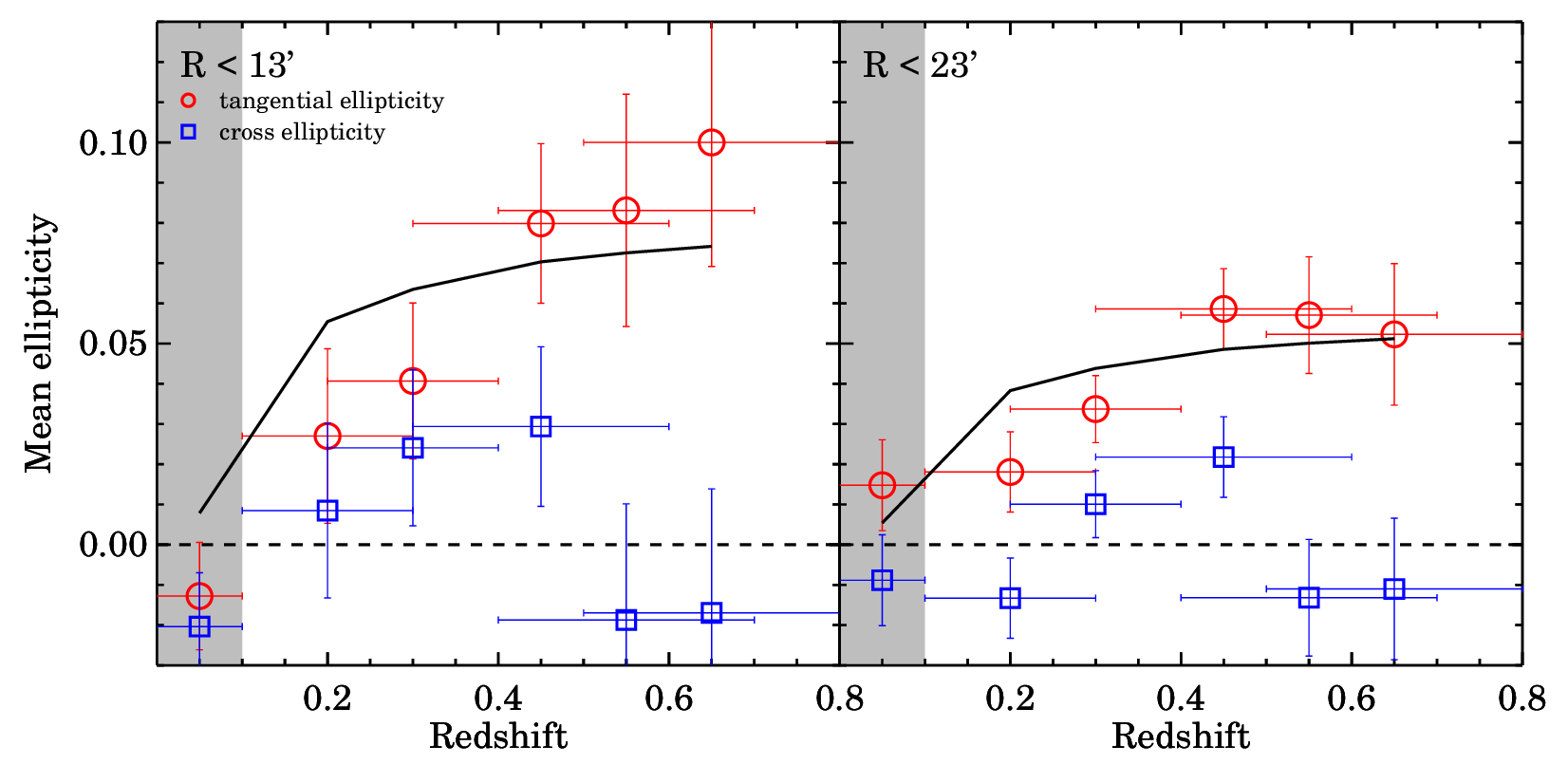}
\caption{Azimuthally averaged mean tangential ellipticity (red circles) and cross ellipticity (blue squares) 
 as a function of spectroscopic redshift  for A2029
 within (Left) $R = 13\arcmin$ and (Right) $R = 23\arcmin$.  
The data point at the lowest redshift is the average signal from the union of the cluster members and foreground galaxies.  
Bins at redshift greater than that of A2029 cover the ranges $0.1<z<0.3$, $0.2<z<0.4$, $0.3<z<0.6$, $0.4<z<0.7$, and $0.5<z<0.8$.
The curve shows the expected signal for a $c=4$ NFW cluster with $M_{200}=9\times 10^{14} M_\odot$, corresponding to the X-ray derived mass from \citet{Walker12, Sohn19a},
with angular diameter distance scaling derived from actual distribution of spectroscopic redshifts within the bin. }.
\label{fig:tomography}
\end{figure*}

Figure \ref{fig:tomography} shows the mean azimuthally averaged tangential (and cross-) ellipticity in different redshift bins. 
The bin for $z < 0.1$ includes the union of the caustic-selected cluster galaxies and foreground galaxies.
The gray background indicates that this point does not enter the shear computation. 

Higher redshift bins cover the ranges 
 $0.1 < z < 0.3$, $0.2 < z < 0.4$, $0.3 < z < 0.6$, $0.4 < z < 0.7$, and $0.5 < z < 0.8$.
We calculate the tomographic signal both for the sample of galaxies
 within a radius of $23\arcmin$ of the center of A2029 (left), 
 and for a radius of $13\arcmin$ (right). 
As noted before,
 $23 \arcmin$ is the approximate virial radius of the cluster \citep{Sohn19a}.  A 
 $13 \arcmin$ radius corresponds to a projected radius of 1 Mpc. 
The choice of the 1 Mpc radius is dictated by a balance 
 between the expected signal amplitude and the number of sources with a redshift.   
 
  The error bars in figure 5 and subsequent figures are statistical error bars constructed from the mean ellipticity $<e>$ of the galaxies in the bin:  $\Delta e = {<e> \over \sqrt{N}}$.
 
\begin{figure}
\centering
\includegraphics[width=0.6\columnwidth]{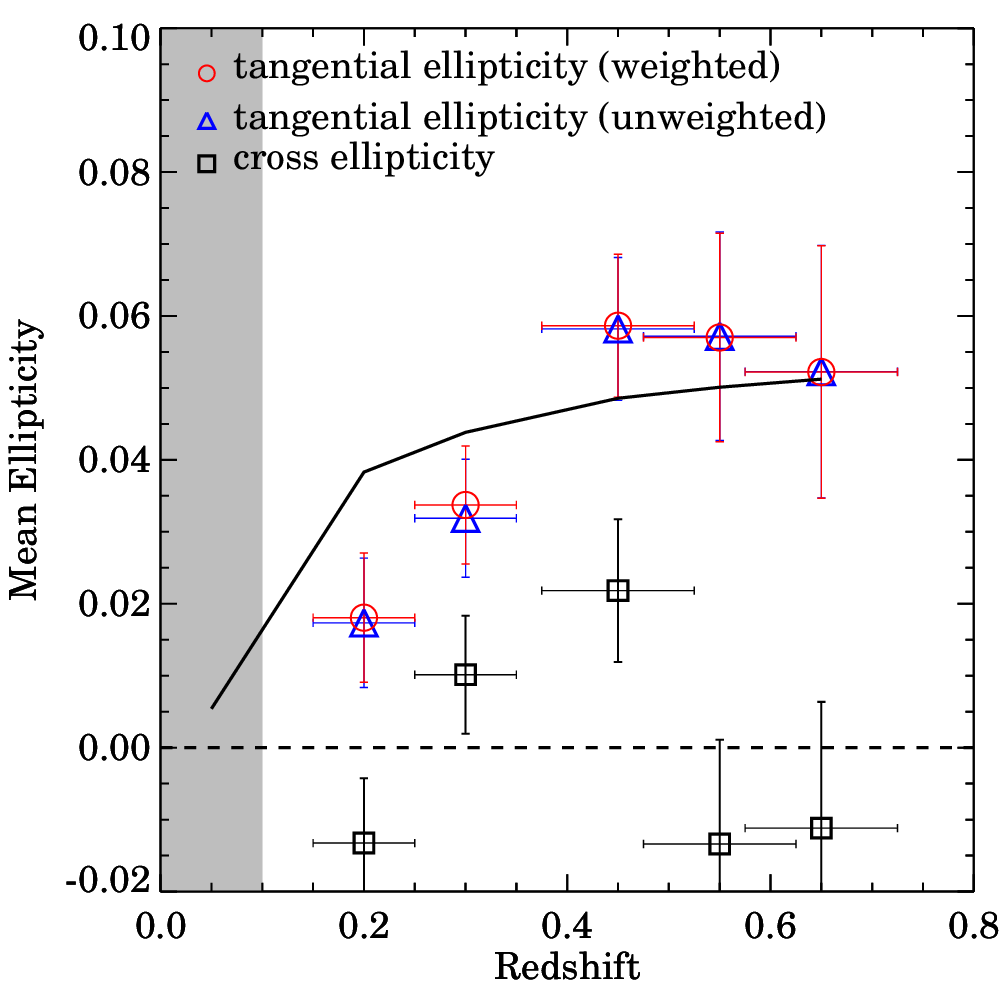}
\caption{The comparison between the spectrotomographic signal using unweighted averages and angular diameter distance ratio corrected averages for the $23 \arcmin$ sample.  The differences are minor and only noticeable for bins with $z<0.4$ because A2029 is a very low redshift. }
\label{weightingeffects}
\end{figure}

  Within each bin, we construct unweighted averages of the tangential ellipticity, rather than weighting by the expected angular diameter distance ratio.  This has the advantage of demonstrating that the signal detected is independent of the weighting.  However, weighting would potentially increase the signal-to-noise for detection in larger spectrotomographic studies, and we would expect that future improvements in the technique would make use of weighting schemes.  In Figure \ref{weightingeffects} we show the difference between calculating the unweighted mean ellipticity or the angular diameter distance weighted in the redshift bins for the 23 arcminute sample.  Because Abell 2029 is at such low redshift, the improvement in sensitivity is noticeable only for the two innermost bins. 
 
For all bins, the B-mode ellipticity is consistent with zero within $\sim 1\sigma$,
 except for the $23\arcmin$ at $0.3<z<0.6$ bin, 
 where it is approximately $2\sigma$ positive.  
This single marginally significant departure is not a serious concern.

We superimpose the expected mean tangential ellipticity 
 (scaled to the mean angular diameter distance of the galaxies) 
 in each bin for a cluster modeled as an NFW with $c=4$ and $M_{200} = 9 \times 10^{14} M_{\odot}$, 
 in accordance with the X-ray derived mass in \citet{Sohn19a}.  
We emphasize that we do not fit the curve to the data points to derive a best-fit NFW model. 
Given the S/N of our detection, the uncertainty would too great.

We calculate the significance of the detection of 
 a shear signal by constructing $10^{6}$ realizations of the shape catalog  
 using randomly oriented galaxies with the distribution of ellipticities in the real catalog.  
Furthermore, we divide the galaxies into 
 different redshift bins as in Figure \ref{fig:tomography}, 
 and repeat the random realizations in each bin.  
Thus, we reject random realizations with large positive signal at low redshift. 
There are 51 (117) realizations exceeding the measured signal within $23\arcmin$ ($13\arcmin$)
 corresponding to $3.9\sigma$ ($3.7\sigma$). 
These values measure the significance of the detection of the lensing signal from A2029. 

Although the $23\arcmin$ sample is slightly more significant,
 the detection significance is not strongly dependent on the area sampled; 
 the same pattern is evident in both samples.  
The amplitude of the mean tangential ellipticity signal for galaxies
 projected closer to the center of A2029 shows a larger shear signal (as expected). 
This result shows the power of the spectroscopy which segregated the global signal of 
 Figure \ref{fig:tomography} as a function of redshift.
This detection is the first tomographic lensing signal for a cluster 
 based exclusively on galaxies with actual spectroscopic redshifts; 
 it is the first spectrotomographic detection.

\begin{figure}
\centering
\includegraphics[width=0.95\columnwidth]{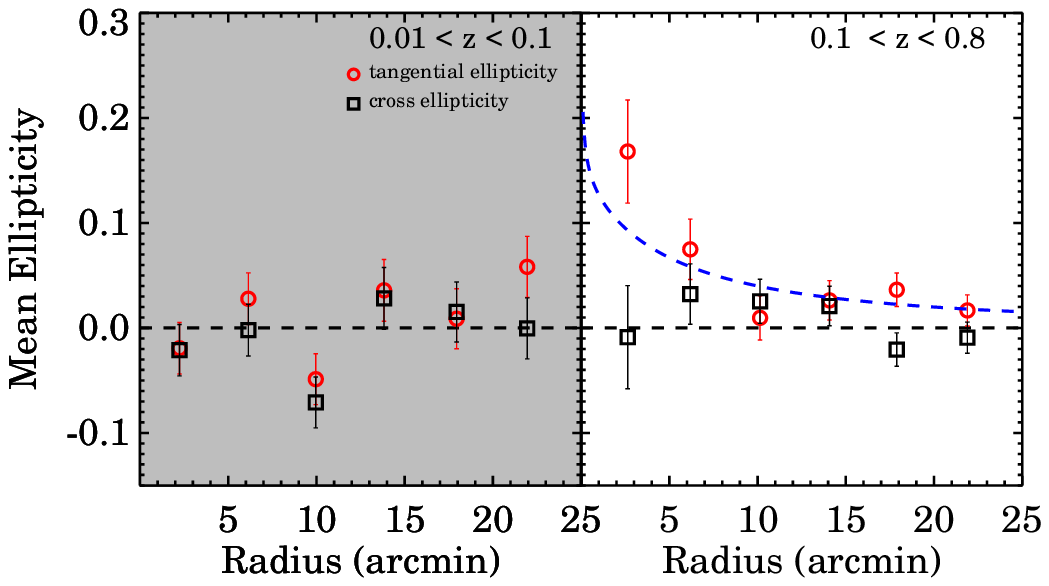}
\caption{Azimuthally averaged tangential ellipticity (and cross ellipticity) 
 as a function of radius from the center of A2029 for galaxies with measured spectroscopic redshifts.  
(Left) The tangential and cross ellipticity profile for galaxies in and in front of the cluster, 
  which should not show any gravitational lensing signal.  
(Right) The tangential and cross ellipticity profile for galaxies with $0.1< z <0.8$.  
The curve is the NFW profile for a cluster with $M=9.0\times 10^{14} M_{\odot}$, $c=4$, 
 scaled to the mean angular diameter distance ratio for the spectroscopic sample.}
\label{fig:tangentialshear_withredshiftg0.3}
\end{figure}

The difference in overall signal level in the left and right panels of Figure \ref{fig:tomography}
 demonstrates that, as expected, 
 there is a measurable radial gradient in the tangential ellipticity.  
We plot the tangential ellipticity versus radius for all galaxies with $z_{spec} > 0.1$ 
 in the right panel of Figure \ref{fig:tangentialshear_withredshiftg0.3}.   
The overplotted curve again corresponds to an NFW profile of tangential ellipticity 
 expected for a cluster with $M_{200}=9.0\times 10^{14} M_{\odot}$ and $c=4$, 
 scaled for the mean angular diameter distance ratio for the sample 
 (a great advantage of a spectroscopic redshift sample is that the distance ratio scaling can be calculated exactly).
The radial profile behaves roughly as expected.
This exercise demonstrates that the tomographic signal 
 we detect is not driven by small-scale features in the distribution of galaxy shapes.
 
The radial ellipticity profiles in Figure \ref{fig:tangentialshear_withredshiftg0.3} 
 also provide additional evidence that the shapes are unbiased.  If there were a radially-dependent measurement bias for the galaxy ellipticities, we would expect it to be fairly independent of the galaxy redshift (because the convergence is small and because the galaxies are well-resolved regardless of their redshift).   
This radially-dependent bias would show up as a constant additive offset 
 in the mean tangential ellipticity at all redshifts.  
The lack of systematic offset in the cluster and low-redshift bins shows that
 the spectroscopic selection does not bias the resulting lensing calculation.
Including only galaxies that are spectroscopically confirmed to be behind the cluster 
 leads to an increased shear signal compared with 
 the signal from the photometric sample of \citep{MDvdL_2018}. 
This behavior again highlights one of the strengths of a spectroscopic sample;
 there is no calibration required to correct for the unknown distribution of 
 angular diameter distance ratios.

\section{Limitations and Future Surveys}
\label{sec:Discussion}

We detect the tomographic weak lensing signal from the massive cluster A2029
  using only spectroscopically confirmed background objects. 
The factor limiting the signal-to-noise of the detection is not the ability to measure shears. 
These  measurements are now relatively easily acquired for galaxies
 behind dozens of galaxy clusters and data for hundreds more will be readily available 
 thanks to ongoing and upcoming surveys with HSC and LSST.  

The main limitation is the  availability of deep, dense redshift surveys.  
We detect the signal for A2029 because it is a very massive cluster, 
 and because its redshift is low enough that a dense redshift survey complete to $r \sim 20.5$ 
 includes enough background galaxy redshifts.  

With the imminent arrival of wide-field spectrographs on large telescopes 
 such as Subaru/PFS \citep{Tamura16} and subsequently of GMACS on the GMT \citep{DePoy18}, 
 much larger redshifts surveys reaching to much fainter magnitudes 
 will be feasible with reasonable telescope time allocations. 
Our experiment demonstrates that 
 these observations will lead to high-significance measurements of individual clusters, 
 opening observational techniques of spectrotomography 
 as a probe of cluster mass distributions and cosmology \citep{vonderLinden14, Hoekstra15, McClintock19}. 

As an example of the power of these observations, 
 we consider galaxy cluster measurements achievable with PFS.
We model the tomographic signals  achievable with a PFS survey complete to $i=22.7$ (corresponding to $\sim 1.5$ hour of exposure per cluster) or $i=23.7$ corresponding to 6 hours/cluster. 
To estimate the number of resolved targets within our $23\arcmin$ radius field, 
 we use the galaxy source density derived from deep, high-quality ($\sim 0.5\arcsec$) 
 seeing imaging of the F2 field \citep{Utsumi16}.  
This photometric sample represents a more typical image quality for future source selection from deep imaging surveys than 
 the A2029 data itself which is taken in moderate seeing. 
We use the measured source counts in F2 to construct two catalogs, 
 one representing a selection complete to $i<22.7$ (6540 galaxies), 
 and one complete to $i<23.7$ (14510 galaxies).

For each galaxy in a catalog, 
 we assign a redshift based on the distribution of photometric redshifts versus magnitude 
 from \citet{Ilbert09}.
Next, we assign the galaxy a cross ellipticity drawn from 
 a zero-centered distribution with a dispersion equal to the dispersion in our measured $e_{1}$,$e_{2}$ distributions for A2029.  
We also assign a tangential ellipticity given by the model for an NFW cluster with $M=9.0\times 10^{14} M_{\odot}$. 
The dispersion of a mean tangential ellipticity is the same as for the cross ellipticity (see Figure \ref{fig:etanecrosshist}).

\begin{figure}
\centering
\includegraphics[width=0.7\columnwidth]{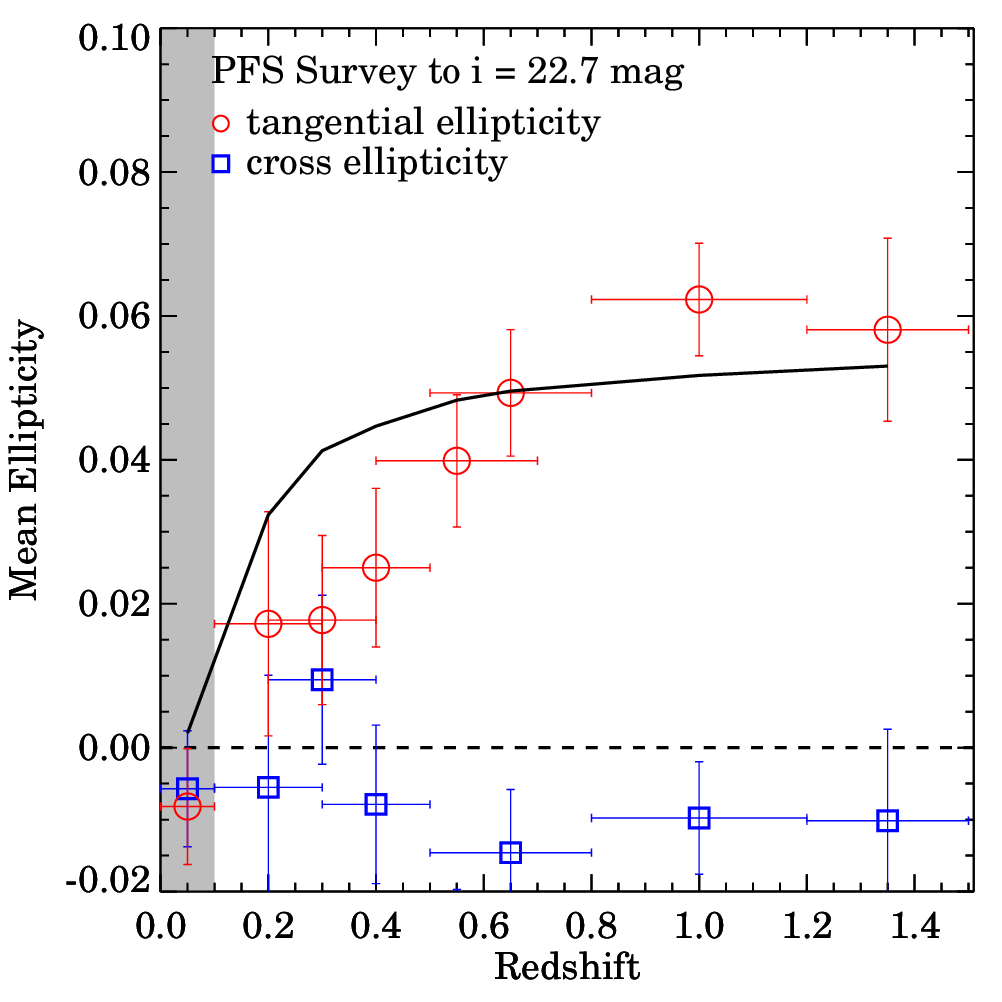}
\caption{Azimuthally averaged tangential (and cross) ellipticity for galaxies as a function of redshift, 
 for a single realization of a model PFS survey. 
The model galaxies are within $i=22.7\arcmin$ of the center of A2029.
The galaxy counts  are based on Subaru HSC $0.5 \arcsec$ imaging \citep{Utsumi18} and 
 reshifts from zCOSMOS \citep{Ilbert09}.  
We assume the mean tangential ellipticity from a pure NFW model for the cluster assuming the dispersion in ellipticities matches that of the e$_{1}$ and e$_{2}$ distributions we observe. }
\label{fig:pfs23}
\end{figure}

\begin{figure}
\centering
\includegraphics[width=0.7\columnwidth]{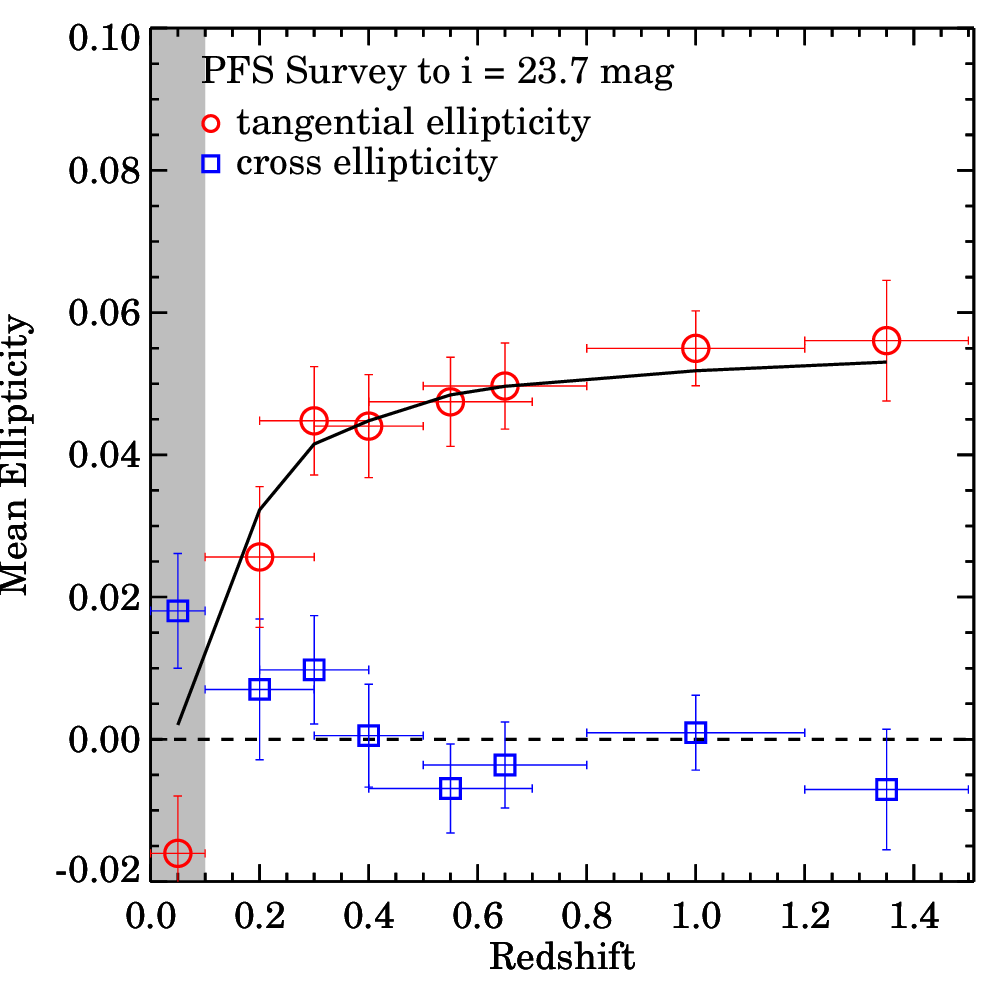} 
\caption{Simulation of a PFS survey complete to $i=23.7$, with the same assumptions as in Figure \ref{fig:pfs23}. }
\label{fig:pfs24}
\end{figure}

We use those distribution of redshifts and simulated ellipticities to construct realizations of tomographic measurements with PFS.  
Figures \ref{fig:pfs23} and \ref{fig:pfs24}, respectively, show realizations of 
 the spectrotomographic measurements for a single realization complete to $i=22.7$ and $i=23.7$.  
For each realization, we calculate the signal-to-noise of the detection of the tomographic signal, 
 extending our calculation to include redshift bins out to $z=1.5$.  

Note that, in addition to the reduction of the error bars in the redshift range explored in this work,
 a PFS survey would extend the redshift distribution of the background sources out to $z \ge 1.5$.
The result is an impressive increase in the detectability of the tomographic signal at high redshift.  

The typical simulated PFS survey to $i<22.7$ reaches a signal-to-noise ratio of 
$\sigma=7.4$; the $i<23.7$ survey reaches $\sigma =11.7$.  
These results are comparable to the significance reached by 
 current photo-z based lensing cluster measurements.  

Carrying out these surveys with PFS would be straightforward, given the 2400-fiber multiplexing.  
Carrying out a survey complete to $i=22.7$ requires $\sim 4$ hours of exposure per cluster.  
A survey complete to $i=23.7$ would obviously be slower; each cluster would take $\sim 2.5$ nights. 
The observations would also be extremely valuable for the study of the star formation and the stellar content cluster galaxies and of the field galaxy populations at intermediate redshift. 
Potentially samples of $\sim 50$ clusters could be observed even to $r=24$, 
 allowing measurement of the slope of the angular diameter distance ratio to 
 $(d{\bf \gamma}/ dz) \sim 0.1$ to a redshift of 1.5.  This would provide a geometric test of the expansion history of the Universe that is independent of the more common tests at these redshifts involving Baryon Acoustic Oscillations (e.g. \citep{2018ApJ...863..110B})) and Type Ia Supernovae (e.g. \citep{2019ApJ...872L..30A}); examining the constraining power of these measurements by themselves or in combination with other techniques will be the goal of subsequent papers.

PFS (and eventually the multi-object spectrographs on giant telescopes) will enable weak lensing measurements 
 based on purely spectroscopic samples of background galaxies with comparable precision to those 
 currently made with photometrically selected samples.  
Larger telescopes will enable these measurements for clusters at larger redshift. 

Although biases in the current photometric approaches can be corrected, in principle, 
 by stacking the signals of large ensembles of clusters \citep{Costanzi19}, 
 large spectroscopic samples, though smaller than the ones suggested here, 
 are necessary to calibrate these biases (which change with cluster redshift and possibly cluster mass).  
The photometric selection of background objects remains subject to foreground contamination.

Spectroscopic redshifts completely eliminate biases resulting from errors in the inferred redshift distributions.  
This approach also eliminates correlations in the signals between tomographic bins originating from sizable errors (random and systematic) 
 in photometric redshifts. 
Furthermore, spectroscopic samples avoid dilution of the signal that arises from the contamination of background shear samples by cluster galaxies. 

Development of spectrotomography will extend the impact of {spectrotomographic} weak lensing from measurement of mass distributions to the determination of cosmological parameters.
Figure \ref{fig:pfs24} shows that constructing deep spectroscopic samples behind even a small sample ($\sim 50$) clusters enables  
 a measurement of the actual shape of the angular diameter distance ratio curve.  
Because this curve depends on the expansion history of the Universe, 
 these samples can provide probes of the cosmological parameters that are independent of other methods \citep{Martinet15}.
In particular, spectrotomography provides a purely geometric test of cosmology, functioning as a completely independent test of cosmography. 

\section{Conclusions}

Weak lensing has a broad impact on astrophysics from the study of cluster masses 
 \citep{Dahle02, Okabe10, Medezinski10, Umetsu14, vonderLinden14, Hoekstra15, McClintock19} and 
 the relationship between the galaxy and dark matter distributions on large scales 
 even including the detection of voids \citep{Davies18}. 
So far the applications have been limited to the use of photometric data. 
For individual clusters the significance of detections has increased steadily from $\sim 4$ \citep{Tyson90} 
 to $\sim 15$ \citep{Okabe10, Schrabback18}. 
Stacking of large samples has further enhanced the impact of the measurements \citep{McClintock19}.

\citet{Hu02} suggested that 
 ultimately spectroscopic tomography could become a powerful astrophysical tool for 
 system mass measurements and, ultimately, for sensitive cosmological tests.
Application of spectroscopic tomography requires large, dense complete redshift surveys to significant depth. 
Until there have been no such surveys in part because of the time on large telescopes required to make the observations.

Here we report a first spectrotomographic detection. 
We use the massive cluster A2029 as a test case.  
This cluster ($z = 0.078$) has a mass of $9\times 10^{14} M_\odot$ \citep{Sohn19a}. 
The low redshift enables an exploration of the spectrotomographic signal with 
 a redshift survey for objects brighter that $r = 20.5$. 
The survey we use contains 1517 background objects within $23\arcmin$ of the cluster weak lensing center.

We detect the spectrotomographic signal at a significance of 
 $3.9\sigma$ within a $23\arcmin$ radius (and 3.7 $\sigma$ within $13 \arcmin$). 
We demonstrate that the signal appears cleanly in the relative distributions of 
 cross and tangential ellipticities for sources with spectroscopic redshifts.
We also demonstrate that these sources have identical distributions of $e_{1}$ and $e_{2}$. 
Other demonstrations of the robustness of the detection included 
 its insensitivity to the exact position of the cluster center and to the redshift binning.

From the physical point of view, 
 the spectrotomographic detection shows the expected dependence of the signal 
 on the source redshift distribution. 
Measurements of this signal with much large samples hold promise for 
 a purely geometric cosmological test.

To explore the promise of spectrotomography we simulate observations with PFS on Subaru.
We show that with complete spectroscopic samples to a limiting $i = 22.7$  and $i = 23.7$ 
 the detections should be in the range $7$ to $12\sigma$.
For the sample to $i = 23.7$, typical surveys would include $\sim 6000$ cluster members 
 and $\sim 15000$ background objects.
Thus ancillary benefits of these surveys include foundations 
 for the spectroscopic study of both cluster and background galaxy populations 
 with unprecedented depth and completeness.

Even for the $i=22.7$ survey case, 
 the error bars in the spectroscopic tomography become small enough to support a cosmological probe. 
The time require for each cluster observation ($\sim 4$ hours/clusters) implies that 
 samples of $>100$ clusters at redshifts similar to A2029 could be observed with
 a reasonable time allocation. 
Samples of this size would open an era of spectroscopic tomography. 
With multi-object spectrographs on larger telescopes such as GMT, 
 these studies could readily be extended to clusters at larger redshift.

\bigskip

\acknowledgments
I.D. acknowledges support from DOE award DE-SC0010010  
J.S. is supported by a CfA Fellowship. 
M.J.G. acknowledges the Smithsonian Institution for support.
This paper includes data produced by the OIR Telescope Data Center in the Smithsonian Astrophysical Observatory.
This research has made use of NASA`s Astrophysics Data System Bibliographic Services. 
Some observations reported here were obtained at the MMT Observatory, a joint facility of the University of Arizona and the Smithsonian Institution.
Some observations reported here were obtained with the Dark Energy Camera installed on the Blanco Telescope at CTIO, part of NSF's National Optical-Infrared Astronomy Research Laboratory.  The National Optical-Infrared Astronomy Research Laboratory is operated by the Association of Universities for Research in Astronomy (AURA) under cooperative agreement with the National Science Foundation.
Data was extracted from the  CSDC Science Data Archive at NSF's National Optical-Infrared Astronomy Research Laboratory 
This research uses the LSST Science Pipelines for shape measurement. Financial support for LSST comes from the National Science Foundation (NSF) through Cooperative Agreement No. 1258333, the Department of Energy (DOE) Office of Science under Contract No. DE-AC02-76SF00515, and private funding raised by the LSST Corporation. The NSF-funded LSST Project Office for construction was established as an operating center under management of the Association of Universities for Research in Astronomy (AURA). The DOE-funded effort to build the LSST camera is managed by the SLAC National Accelerator Laboratory (SLAC).

\bibliographystyle{apj}
\bibliography{A2029wlz_apjMJS_0610}

\end{document}